\documentclass[12pt]{article}
\usepackage{graphicx}  
\DeclareGraphicsExtensions{.eps.gz,.eps,.ps,.ps.gz}
\parskip=\itemsep               
\date{\empty}
\title{
{\normalsize\rightline{DESY 99-136}\rightline{hep-ph/9909338}} 
\vskip 1cm 
{\bf Theory and Phenomenology of Instantons at HERA}\thanks{
Contribution to the Ringberg Workshop ``New Trends in HERA Physics'',    
Ringberg Castle, Tegernsee, Germany, May 30 - June 4, 1999; to be
published in the Proceedings.} 
       \vspace{15mm}}
\author{A. Ringwald and F. Schrempp\\[4mm] 
Deutsches Elektronen-Synchrotron DESY, Hamburg, Germany\\[20mm]}

\begin{document}

\begin{titlepage}
\maketitle              
\begin{abstract}
We review our on-going theoretical and phenomenological investigation
of the pros\-pects to discover QCD-instantons in deep-inelastic
scattering at HERA. 
\end{abstract}
\thispagestyle{empty}
\end{titlepage}
\newpage \setcounter{page}{2}

\section{Introduction}

It is a remarkable fact that non-Abelian gauge fields in four 
Euclidean space-time dimensions carry an integer topological charge. 
Instantons~\cite{bpst} (anti-instantons) are classical solutions of the 
Euclidean Yang-Mills equations and also represent the simplest 
non-perturbative 
fluctuations of gauge fields with topological charge $+1$ ($-1$). 
In QCD, instantons are widely believed to play an essential r{\^o}le
at long distance: They provide a solution of the axial $U(1)$ 
problem~\cite{th}, 
and there seems to be some evidence that they induce
chiral symmetry breaking and affect the light hadron 
spectrum~\cite{ssh}. Nevertheless, a direct experimental observation of
instanton-induced effects is still lacking up to now. 

Deep-inelastic scattering at HERA offers a unique window to 
discover QCD-instanton induced events directly through their characteristic
final-state signature~\cite{rs,grs,jgerigk,cgrs} and a sizeable rate, 
calculable within instanton perturbation theory~\cite{mrs1,rs-pl,mrs2}. 
It is the purpose of the present contribution to review our
theoretical and phenomenological investigation of the prospects
to trace QCD-instantons at HERA.  

The outline of this review is as follows:

We start in Sect.~\ref{I-vacuum} with a short introduction
to instanton physics, contentrating especially on two important
building blocks of instanton perturbation theory, namely the 
instanton size distribution and the instanton-anti-instanton 
interaction. 
A recent comparison~\cite{rs-lat} of the 
perturbative predictions of these quantities with their non-perturbative 
measurements on the lattice~\cite{ukqcd} is emphasized. It allows to extract 
important information about the range of validity of instanton perturbation 
theory. 
The special r{\^o}le of deep-inelastic scattering in instanton physics 
is outlined in Sect.~\ref{I-dis}: The Bjorken variables of instanton
induced hard scattering processes probe the instanton size distribution
and the instanton-anti-instanton interaction~\cite{mrs1,rs-pl}. By final
state cuts in these variables it is therefore possible to stay within the 
region of applicability of instanton perturbation theory, inferred from our
comparison with the lattice above. Moreover, within this fiducial kinematical
region, one is able to predict the rate and the (partonic) final state.
We discuss the properties of the latter as inferred from our
Monte Carlo generator QCDINS~\cite{grs,rs-qcdins}.
In Sect.~\ref{search}, we report on a possible search strategy for 
instanton-induced processes in deep-inelastic scattering at HERA~\cite{cgrs}. 

\section{\label{I-vacuum}Instantons in the QCD Vacuum}

In this section let us start with a short introduction to instantons
and their properties, both in the perturbative as well as in the
non-perturbative regime. We shall concentrate on those aspects that will
be important for the description of instanton-induced scattering processes in 
deep-inelastic scattering in Sect.~\ref{I-dis}. In particular, we shall 
report on our recent determination of the region of applicability of instanton
perturbation theory for the instanton size distribution and the
instanton-anti-instanton interaction~\cite{rs-lat}.  
Furthermore, we elucidate the connection of instantons with the
axial anomaly.

Instantons~\cite{bpst}, being solutions of the Yang-Mills equations in
Euclidean space, are minima of the Euclidean action $S$. 
Therefore, they appear naturally as generalized saddle-points in the Euclidean 
path integral formulation of QCD, according to which the 
expectation value of an observable ${\mathcal O}$ is given by
\begin{eqnarray}
        \langle \mathcal{O}[A,\psi,\overline{\psi}]\rangle 
        =\frac{1}{Z} \int [dA][d\psi][d\overline{\psi}]\, 
        \mathcal{O}[A,\psi,\overline{\psi}]\,
        {\rm e}^{-
        S[A,\psi,\overline{\psi}]
                }\, , 
\label{eucl-path}
\end{eqnarray}
where the normalization, 
\begin{eqnarray}
        Z=\int [dA][d\psi][d\overline{\psi}] \,
        {\rm e}^{-
        S[A,\psi,\overline{\psi}]
                } \, ,
\label{part}
\end{eqnarray}
denotes the partition function.
Physical observables (e.g. $S$-matrix elements) are obtained from the Euclidean
expectation values~(\ref{eucl-path}) by analytical continuation 
to Min\-kow\-ski space-time. In particular, the partition 
function~(\ref{part}) corresponds 
physically to the vacuum-to-vacuum amplitude.

Instanton perturbation theory results from the generalized saddle-point
expansion of the path integral~(\ref{eucl-path}) about 
non-trivial minima of the Euclidean action\footnote{Perturbative QCD is 
obtained from an expansion about the perturbative vacuum solution, i.e. 
vanishing gluon field and vanishing quark fields and thus vanishing 
Euclidean action.}. It can be shown that these non-trivial solutions have
integer topological charge,
\begin{eqnarray}
      { Q} \equiv \frac{\alpha_s}{2\pi} 
                   \int d^4x\, \frac{1}{2}{\rm tr}
                   (F_{\mu\nu}{\tilde F}_{\mu\nu}) 
              = { \pm 1}, { \pm 2}, \ldots ,
\end{eqnarray}
and that their action is a multiple of $2\pi/\alpha_s$,
\begin{eqnarray}
      S\equiv  \int d^4x\, \frac{1}{2}{\rm tr}
                   (F_{\mu\nu}F_{\mu\nu}) = 
                 \frac{2\,\pi}{\alpha_s}\,| Q |
          = \frac{2\,\pi}{\alpha_s}\cdot 
      ({ 1},{ 2}, \ldots) .
\end{eqnarray}
In the weak coupling regime, $\alpha_s\ll 1$, the dominant saddle-point has 
$|Q|=1$. The solution corresponding to $Q=1$ is given by\footnote{In 
Eq.~(\ref{instanton}) and throughout the paper 
we use the abbreviations,
$v \equiv v_\mu \sigma^\mu$,  $\overline{v} \equiv v_\mu 
\overline{\sigma}^\mu$ for any four-vector $v_\mu $.}~\cite{bpst} 
(singular gauge)\begin{eqnarray}
      A_{\mu}^{{ (I)}}(x; \rho , U, x_0)=
      -\frac{\rm i}{ g}\,
      \frac{{ \rho}^{2}}{(x-{ x_0})^2}
      \,{ U}\,\frac{
      \sigma_{\mu}\,(\overline{x}-\overline{{ x_0}})-
      (x_{\mu}-{ x_0}_\mu)}
      {(x-{ x_0})^2+{ \rho}^2}\,
      { U}^{\dagger} , 
\label{instanton}
\end{eqnarray}
where the ``collective coordinates''  $\rho$, $x_0$ and $U$  denote the size,
position and colour orientation of the solution. 
The solution~(\ref{instanton}) has been called ``instanton'' ($I$), since it 
is localized in Euclidean space 
and time (``instantaneous''), as can be seen from its Lagrange density, 
\begin{eqnarray}
\mathcal{L}\left(A_\mu^{{ (I)}}(x;\rho,U,x_0)\right)
          =      \frac{12}{\pi\alpha_s}\cdot\frac{\rho^4}
        {((x-x_0)^2+\rho^2)^4}\ 
      \Rightarrow S\left[A_\mu^{{ (I)}}\right] =
      \,\frac{2\,\pi}{\alpha_s}\, . 
\label{lagr}
\end{eqnarray}
It appears as a spherical symmetric bump of size $\rho$ centred at $x_0$. 
\begin{figure}
\begin{center}
\includegraphics[width=.3\textwidth]{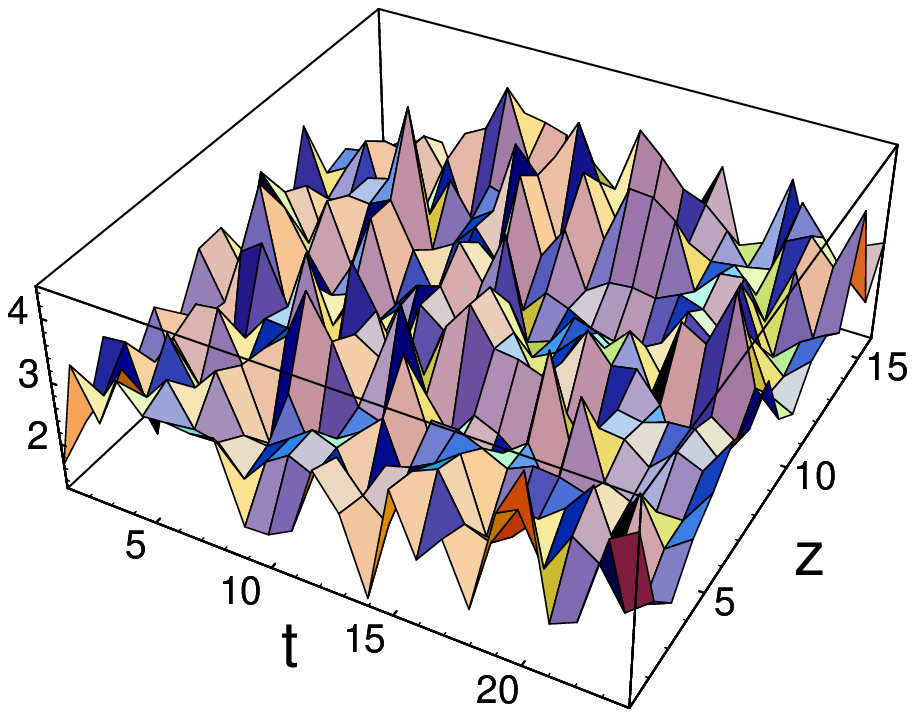}
\includegraphics[width=.3\textwidth]{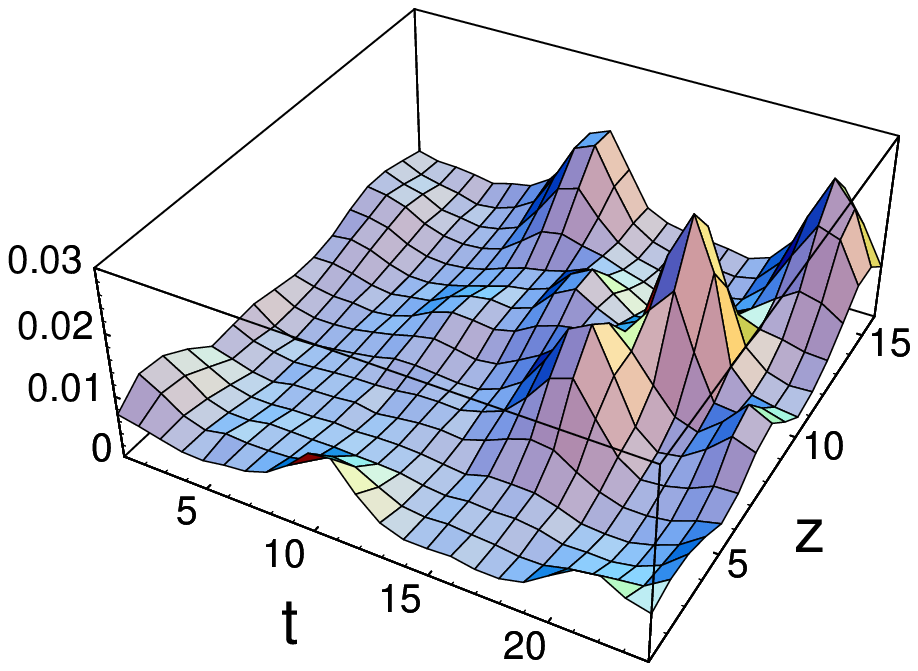}
\hspace{0.05\textwidth}
\includegraphics[width=.3\textwidth]{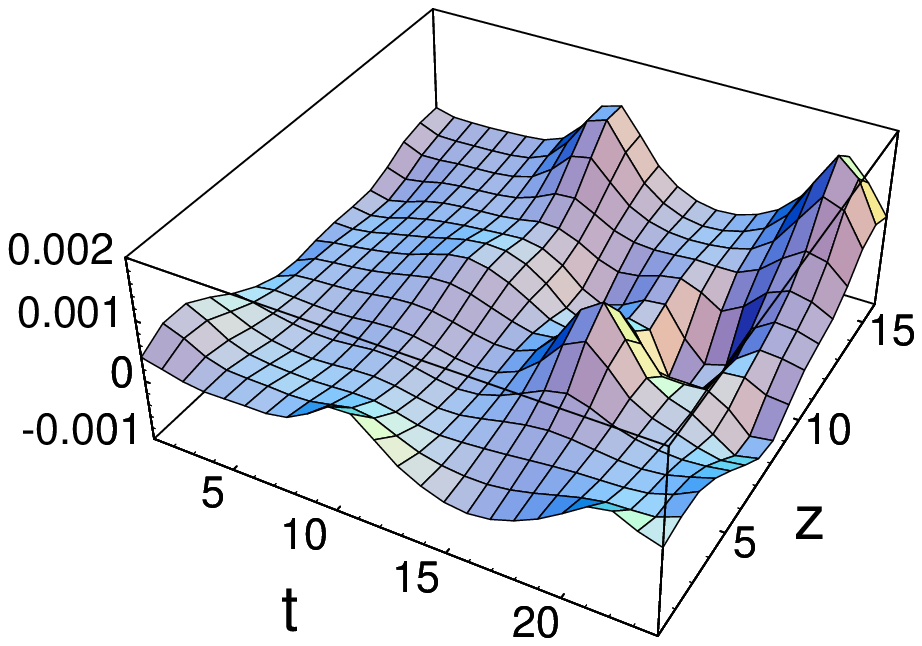}
\put(-150,120){\small\bf Topological Charge Density}
\put(-350,120){\small\bf Lagrange Density}
\end{center}
\caption[]{Instanton content of a typical
slice of a gluon configuration on the lattice at fixed $x$, $y$ as a function 
of $z$ and
$t$~\cite{chu}. Lagrange density before ``cooling'', with  fluctuations of 
short wavelength ${\cal O}(a)$ dominating (left). After ``cooling'' by 25 
steps, 3 $I$'s and 2 $\overline{I}$'s may be clearly identified as bumps in 
the Lagrange density (middle) and the topological charge density (right).}
\label{lattice-intro}
\end{figure}

The natural starting point of instanton perturbation theory is the 
evaluation of the instanton contribution to the partition 
function~(\ref{part})~\cite{th}, by expanding the path integral 
about the instanton~(\ref{instanton}). Since the action is independent
of the collective coordinates, one has to integrate over them 
and obtains the $I$-contribution $Z^{(I)}$, normalized to the topologically 
trivial perturbative contribution $Z^{(0)}$,
in the form\footnote{For notational
simplicity, we call the $I$-position in the following $x$ (instead of $x_0$). 
}    
\begin{eqnarray}
  \frac{1}{Z^{(0)}} \frac{d Z^{(I)}}{d^4x}=
      \int\limits_0^\infty d\rho\, D_m (\rho )\int dU \,.
\label{zi}
\end{eqnarray}
The size distribution $D_m (\rho )$ is known in the 
framework of $I$-perturbation theory for small 
$\alpha_s(\mu_r)\ln(\rho\,\mu_r)$ and small $\rho\,m_i(\mu_r)$, where 
$m_i(\mu_r)$ are the running quark masses and $\mu_r$ denotes the 
renormalization scale. 
After its pioneering evaluation at 
1-loop~\cite{th}  for $N_c=2$ and its generalization~\cite{ber} to
arbitrary  $N_c$, it is meanwhile available~\cite{morretal} in 2-loop
renormalization-group (RG) invariant form,
i.e. $D^{-1}\,dD/d\ln(\mu_r)=\mathcal{O}(\alpha_s^2)$,
\begin{eqnarray}
\frac{dn_I}{d^4x\,d\rho}=\,
D_m({\rho})= D({\rho})\,\prod_{i=1}^{n_f}(\rho\,m_i(\mu_r)) \,
(\rho\,
\mu_r)^{n_f\,\gamma_0
\frac{\alpha_{\overline{\rm MS}}(\mu_r)}{4\pi}} ,
\label{densm} 
\end{eqnarray}
with the reduced size distribution
\begin{eqnarray}
D({\rho}) 
=
\frac{d_{\overline{\rm MS}}}{\rho^5}
\left(\frac{2\pi}{\alpha_{\overline{\rm MS}}(\mu_r)}\right)^{2\,N_c} 
\exp{\left(-\frac{2\pi}{\alpha_{\overline{\rm MS}}(\mu_r)}\right)}(\rho\,
\mu_r)^{\beta_0+(\beta_1-4\,N_c\beta_0 )
\frac{\alpha_{\overline{\rm MS}}(\mu_r)}{4\pi}} .
\label{dens}
\end{eqnarray} 
Here, $\gamma_0$ is the leading anomalous dimension coefficient, 
$\beta_i$ ($i=0,1$) denote the leading and next-to-leading 
$\beta$-function coefficients and $d_{\overline{\rm MS}}$ is a 
known~\cite{dMS} constant. 
\begin{figure}
\vspace{-3ex}
\begin{center}
\includegraphics[width=0.48\textwidth]{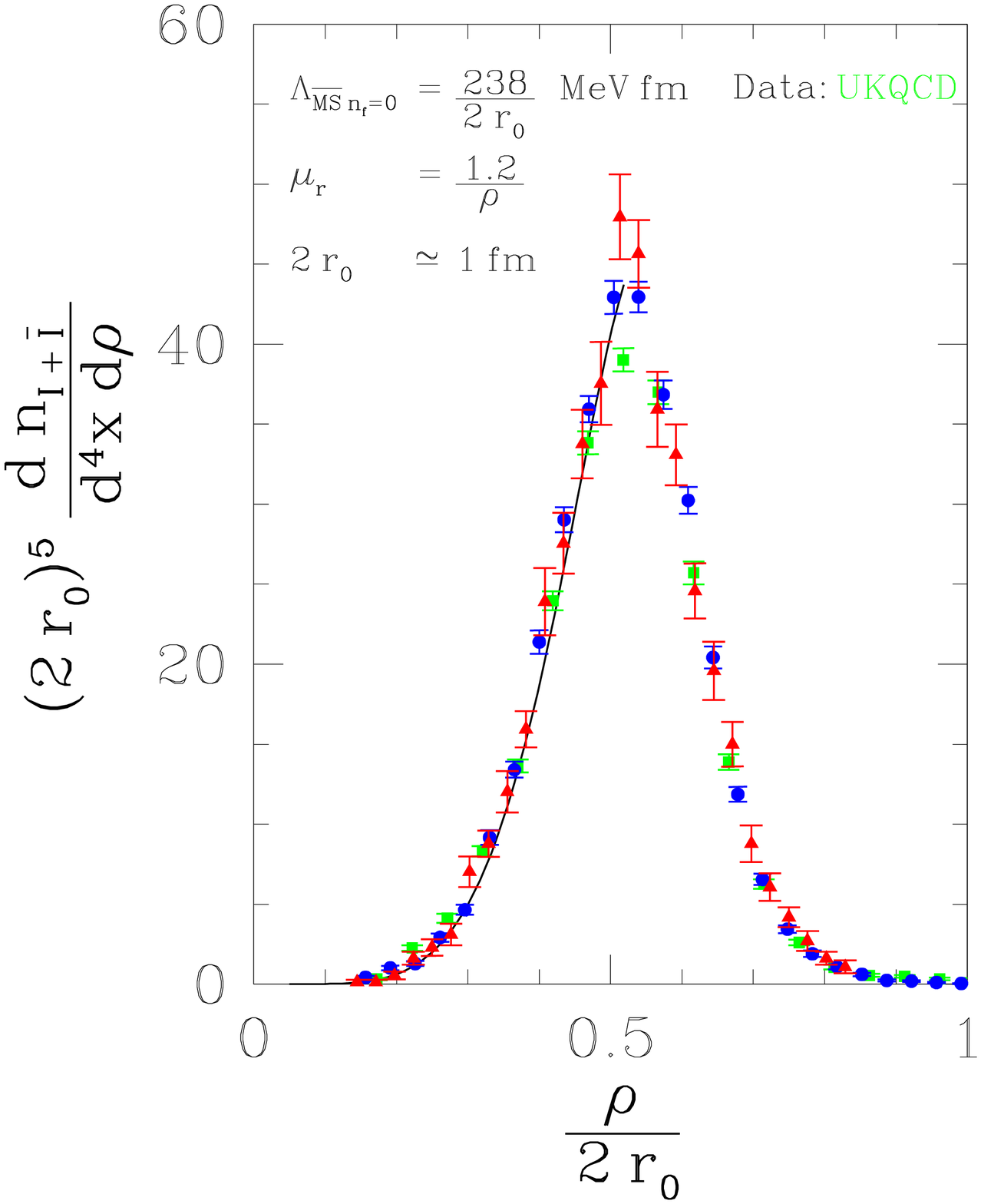}
\vspace{1.5ex}

\includegraphics[width=0.41\textwidth]{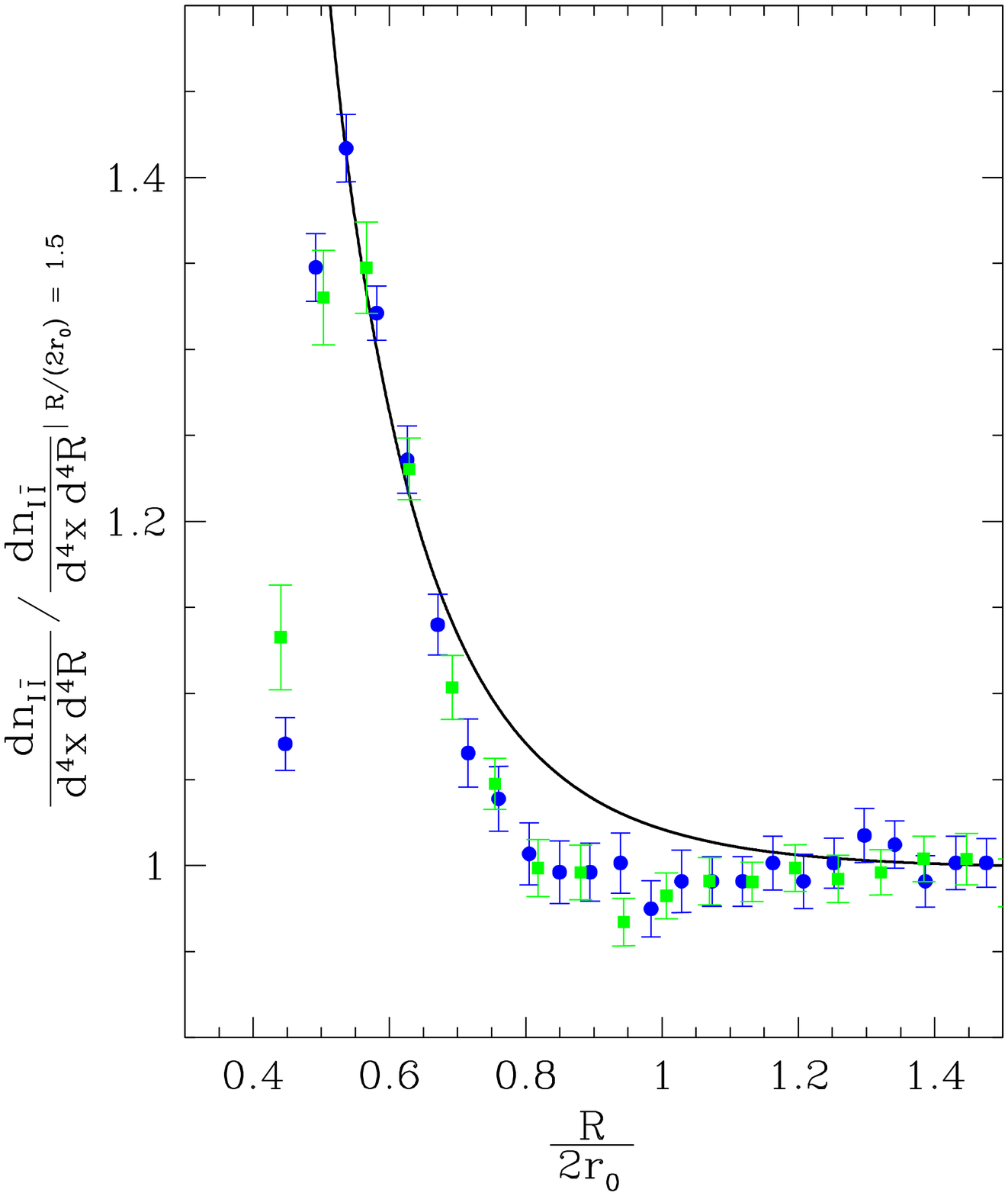}
\end{center}
\caption[]{Continuum limit~\cite{rs-lat} of ``equivalent'' UKQCD
data~\cite{ukqcd,mike} for the ($I+\overline{I}$)-size distribution (top)  
and the normalized $I\overline{I}$-distance distribution (bottom)
along with the 
respective predictions from $I$-perturbation theory and the valley
form of the $I\overline{I}$-interaction~\cite{rs-lat}. The 3-loop form of
$\alpha_{\rm \overline{MS}}$ with $\Lambda_{{\rm
\overline{MS}}}^{(0)}$ from ALPHA~\cite{alpha} was used.  
} 
\label{lattcomp}
\end{figure}

The powerlaw behaviour of the (reduced) $I$-size distribution,
\begin{equation} 
\label{powerlaw}
D({\rho})\sim \rho^{\beta_0 -5 +\mathcal{O}(\alpha_s)},
\end{equation}
generically causes the dominant contributions to the $I$-size
integrals (e.g. Eq.~(\ref{zi})) to originate from the infrared (IR) regime 
(large $\rho$) and thus often spoils the applicability of $I$-perturbation 
theory. Since the $I$-size distribution not only appears in the 
vacuum-to-vacuum amplitude~(\ref{zi}), but also in generic instanton-induced 
scattering amplitudes (c.f. Sect.~\ref{I-dis}) and matrix elements, it is 
extremely important to know the region of validity of the perturbative result 
(\ref{dens}).  

Crucial information on the range of validity comes~\cite{rs-lat} from a 
recent high-quality lattice investigation~\cite{ukqcd} on the
topological structure of the QCD vacuum (for $n_f=0$).
In order to make $I$-effects visible in lattice simulations with
given lattice spacing $a$, the raw data have to be ``cooled'' first.
This procedure is designed to filter out
(dominating) fluctuations of short wavelength ${\cal O}(a)$,
while affecting the topological fluctuations of much longer wavelength 
$\rho \gg a$ comparatively little. After cooling, an ensemble of $I$'s and
$\overline{I}$'s can clearly be seen (and studied) as bumps in the
Lagrange density and 
in the topological charge density (c.f. Fig.~\ref{lattice-intro}).  

Figure~\ref{lattcomp}\,(top) illustrates the striking agreement in shape and
normalization~\cite{rs-lat} of $2\,D(\rho)$ with the continuum limit of the
UKQCD lattice data~\cite{ukqcd} for
$d n_{I+\overline{I}}/d^4x\,d\rho$, for 
$\rho{\mbox{\,\raisebox{.3ex}
    {$<$}$\!\!\!\!\!$\raisebox{-.9ex}{$\sim$}\,}} 0.3-0.35$ fm.   
The predicted normalization of $D(\rho)$ is very sensitive to 
$\Lambda^{(0)}_{{\overline{\rm MS}}}$ for which we took the most accurate
(non-perturbative) result from ALPHA~\cite{alpha}. The theoretically favoured
choice $\mu_r\rho={\cal O}(1)$ in Fig.~\ref{lattcomp}\,(top)
optimizes the range of agreement, extending right up to the peak
around $\rho\simeq 0.5$ fm. However, due to its two-loop
renormalization-group invariance, $D(\rho)$ is almost independent
of $\mu_r$ for $\rho {\mbox{\,\raisebox{.3ex}
    {$<$}$\!\!\!\!\!$\raisebox{-.9ex}{$\sim$}\,}} 0.3$ fm over a wide 
$\mu_r$ range. 
Hence, for 
$\rho {\mbox{\,\raisebox{.3ex}
    {$<$}$\!\!\!\!\!$\raisebox{-.9ex}{$\sim$}\,}} 0.3$ fm, there is
effectively no free parameter involved.

\begin{figure}
\begin{center}
\parbox{.3\textwidth}
{\includegraphics[width=.25\textwidth]{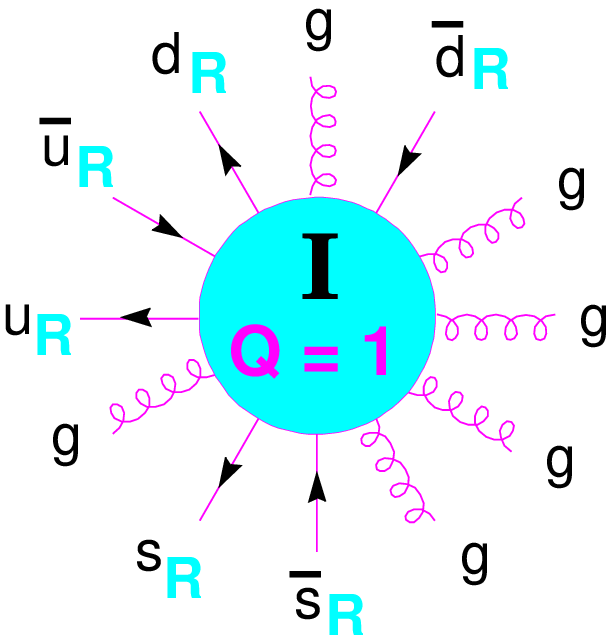}}
\hspace{.05\textwidth}
{$ \Delta\,{ Q_5}= 2\, n_f \cdot { Q}$}
\hspace{.09\textwidth}
\parbox{.3\textwidth}
{\includegraphics[width=.25\textwidth]{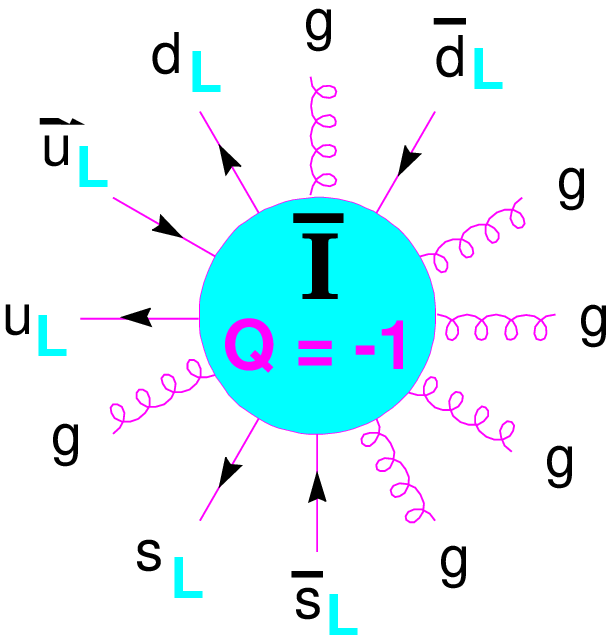}}
\end{center}
\caption[]{Instantons and anti-instantons induce chirality violating 
amplitudes.}
\label{instproc}
\end{figure}

Turning back to the perturbative size distribution~(\ref{densm}) in 
QCD with $n_f\neq 0$ light quark flavours, we would like 
to comment on the appearent suppression of the instanton-induced
vacuum-to-vacuum amplitude~(\ref{zi}) for small quark masses, $\rho m_i\ll 1$. 
It is related~\cite{th} to the axial 
anomaly~\cite{anomaly} according to which 
any gauge field fluctuation with topological charge $Q$ must be accompanied
by a corresponding change in chirality, 
\begin{equation}
\triangle {Q_{5\,i}}=2\,{Q}\,; \hspace{6ex} i=1,\ldots,n_f\,.
\label{anomaly}
\end{equation}
Thus, pure vacuum-to-vacuum transitions induced by 
instantons are expected to be rare. On the other hand, scattering amplitudes 
or Green's functions corresponding to anomalous chirality violation 
(c.f. Fig.~\ref{instproc}) are expected to receive their main contribution 
due to instantons and do not suffer from any mass suppression.      

Let us illustrate this by the simplest example of one light flavour 
($n_f=1$): 
The instanton contribution to the fermionic two-point function can be written 
as
\begin{eqnarray}
        \langle \psi (x_1)\overline{\psi}(x_2)\rangle^{{(I)}} 
        \simeq \int d^4x\int\limits_0^\infty d\rho\,D(\rho )\int dU\, 
        (\rho\, m)\, S^{(I)}(x_1,x_2;x,\rho ,U ) \,.
\label{twopoint}
\end{eqnarray}
Expressing the quark propagator in the $I$-background, $S^{(I)}$, 
in terms of the spectrum of the Dirac operator in the $I$-background,
which has exactly one right-handed zero mode\footnote{According to 
an index theorem~\cite{atiyah}, the number $n_{R/L}$ of 
right/left-handed zero modes of the Dirac operator in the background
of a gauge field with topological charge $Q$ satisfies 
$n_R-n_L=Q$. For the instanton: $n_R=Q=1$; $n_L=0$.}
 $\kappa_0$~\cite{th},
\begin{eqnarray}
        -{\rm i}\not{D}^{{(I)}}\kappa_n&=&\lambda_n\kappa_n;\hspace{8ex}   
     {\rm with}\    {\lambda_0=0}\ {\rm and}\ 
\lambda_n\not=0\ {\rm for}\ n\not=0 , \\
        S^{(I)}(x_1,x_2;\ldots )&=&
        \frac{{\kappa_0}(x_1;\ldots )\,\kappa_0^\dagger (x_2;\ldots )}{{m}} +
        \sum_{n\not= 0} 
     \frac{\kappa_n(x_1;\ldots )\,\kappa_n^\dagger 
(x_2;\ldots )}{m+i\lambda_n},
\end{eqnarray}
we see that for $m\to 0$ only the 
zero mode contribution survives in Eq.~(\ref{twopoint}), 
\begin{eqnarray}
        \langle \psi (x_1)\overline{\psi}(x_2)\rangle^{{(I)}} 
        \simeq \int d^4x\int\limits_0^\infty d\rho\,D(\rho )\int dU
        \rho\, {\kappa_0}(x_1;x,\rho ,U)\,\kappa_0^\dagger (x_2;x,\rho ,U)   .
\label{twopointan}
\end{eqnarray}
Note that $\kappa_0\kappa_0^\dagger$ has ${Q_5}=2$, exactly as required
by the anomaly~(\ref{anomaly}). For the realistic case of three light 
flavours ($n_f=3$), the generalization of
Eq.~(\ref{twopointan}) leads to non-vanishing, chirality violating six-point 
functions corresponding to the anomalous processes shown in
Fig.~\ref{instproc}.

\begin{figure}
\begin{center}
\includegraphics[width=.65\textwidth]{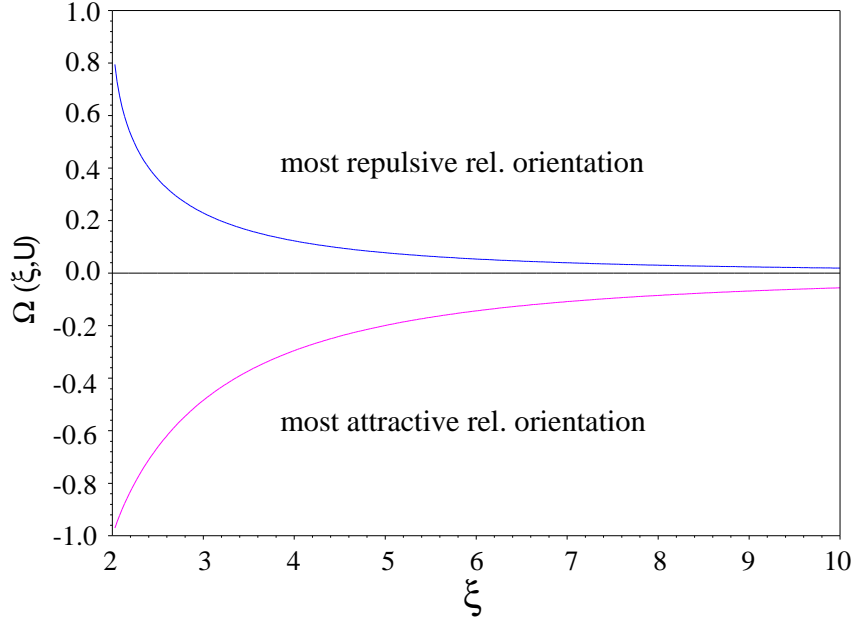}
\end{center}
\caption[]{The instanton-anti-instanton interaction as a function of 
the conformal separation $\xi$, for the most attractive and the 
most repulsive relative orientation, respectively.
          }   
\label{omega}
\end{figure}

Finally, let us turn to the interaction between instantons and 
anti-instantons. In the instanton-anti-instanton ($I\overline{I}$)
valley approach~\cite{yung} it is determined in the following way: 
Starting from the infinitely separated ($R\to\infty$) $I\overline{I}$-pair,  
\begin{eqnarray}
        A_\mu^{{(I\overline{I})}}(x;\rho ,\overline{\rho},U,R)
        \stackrel{{R\to\infty}}{=}
        A_\mu^{{(I)}}(x;\rho,{\bf 1})+
        A_\mu^{{(\overline{I})}}(x-R;\overline{\rho},U) 
\label{sum-ans}
\end{eqnarray}  
one looks for a constraint solution, which is the minimum of the action for 
fixed collective coordinates, $\rho,\overline{\rho},U,R$. 
The valley equations have meanwhile been solved for 
arbitrary separation $R$~\cite{valley-most-attr-orient} and arbitrary relative 
color orientation $U$~\cite{valley-gen-orient}. Due to classical  
conformal invariance, the $I\overline{I}$-action $S^{(I\overline{I})}$ and
the interaction $\Omega$, 
\begin{eqnarray}
      S[A_\mu^{{(I\overline{I})}}]=\frac{4\pi}{\alpha_s}\,
      S^{(I\overline{I})}({\xi},U)=
      \frac{4\pi}{\alpha_s}\,\left( 1+\Omega({\xi},U)
                             \right)
\end{eqnarray}      
depend on the sizes and the separation only through the 
``conformal separation'',
\begin{equation}
\xi =
\frac{R^2}{\rho\overline{\rho}}+\frac{\overline{\rho}}{\rho}+
\frac{\rho}{\overline{\rho}} \, .
\end{equation}
Because of the smaller action, the most attractive relative orientation 
(c.f. Fig.~\ref{omega}) dominates in the weak coupling regime. Thus, in this
regime, nothing prevents instantons and 
anti-instantons from approaching each other and annihilating.

From a perturbative expansion of the path integral about
the $I\overline{I}$-valley, one obtains  
the contribution of the $I\overline{I}$-valley to the partition 
function~(\ref{part}) in the form
\begin{eqnarray}
   \frac{1}{Z^{(0)}} \frac{d Z^{(I\overline{I})}}{d^4x}=
      \int  d^4 R \int\limits_0^\infty d\rho\,
      \int\limits_0^\infty d\overline{\rho}\,
      D_{I\overline{I}}(R,\rho ,\overline{\rho} ) \, ,
\end{eqnarray}
where the group-averaged distribution of $I\overline{I}$-pairs, 
$D_{I\overline{I}}(R,\rho ,\overline{\rho} )$, 
is known, for small $\alpha_s$, $m_i$, and for 
sufficiently large $R$~\cite{rs-pl,mrs2,rs-lat},
\begin{eqnarray}
\label{IaI}
\lefteqn{
\frac{d n_{I\overline{I}}}{d^4 x\,d^4
R\,d\rho\,d\overline{\rho}} \simeq 
      D_{I\overline{I}}(R,\rho,\overline{\rho})=} 
\\ \nonumber &&  
D(\rho )\,
      D(\overline{\rho})
      \ \int dU\, 
      \exp\left[-\frac{4\pi}{\alpha_{\overline{\rm
      MS}}(s_{I\overline{I}}/\,\sqrt{\rho\overline{\rho}})}\, 
 \Omega
 \left(\frac{R^2}{\rho\overline{\rho}},\frac{\overline{\rho}}{\rho},U 
  \right) \right]
      \omega ({\xi},U)^{2\,n_f}
\, . 
\end{eqnarray}
Here, the scale factor $s_{I\overline{I}}=\mathcal{O}(1)$ parametrizes the
residual scheme dependence and   
\begin{eqnarray}
      \omega =\int d^4x\, 
      \kappa_{0\,{I}}^\dagger (x;\ldots )\,[{\rm i}\not{D}^{(I\overline{I})}]\,
      \kappa_{0\,{\overline{I}}}(x-R;\ldots )
\end{eqnarray}
denotes the fermionic interaction induced by the quark zero modes. 

We will see below in Sect.~\ref{I-dis} that the distribution~(\ref{IaI}) 
is a crucial input for instanton-induced scattering cross sections.
Thus, it is extremely welcome that the range of validity of (\ref{IaI}) can 
be inferred from a comparison with recent lattice data.  
Fig.~\ref{lattcomp}\,(bottom) displays the continuum
limit~\cite{rs-lat} of the UKQCD data~\cite{ukqcd,mike} for the
 distance distribution of $I\overline{I}$-pairs,
$dn_{I\overline{I}}/d^4x\,d^4R$,  
along with the theoretical prediction~\cite{rs-lat}. The latter
involves (numerical) integrations of $\exp
(-4\pi/\alpha_s\cdot\Omega)$ over the
$I\overline{I}$ relative color orientation $(U)$, as well as $\rho$ and
$\overline{\rho}$. For the respective weight $D(\rho)D(\overline{\rho})$,  
a Gaussian fit to the lattice data was used in order to avoid convergence
problems at large $\rho,\overline{\rho}$. We note a good
agreement with the lattice data down to $I\overline{I}$-distances
$R/\langle\rho\rangle\simeq 1$. These results imply first direct
support for the validity of the ``valley''-form of the interaction
$\Omega$ between $I\overline{I}$-pairs. 
 
In summary: The striking agreement of the UKQCD lattice data with
$I$-perturbation theory is a very interesting result by itself. The
extracted lattice constraints on the range of validity of
$I$-perturbation theory can be directly translated into a ``fiducial''
kinematical region for our predictions~\cite{rs-pl,rs-lat} in deep-inelastic
scattering, 
as shall be discussed in the next section.

\section{\label{I-dis}Instantons in Deep-Inelastic Scattering}

In this section we shall elucidate the special r{\^o}le of deep-inelastic
scattering for instanton physics. We shall outline that only small 
size instantons, which are theoretically under contr{\^o}l, are probed
in deep-inelastic scattering~\cite{mrs1}. Furthermore, we shall show that 
suitable cuts in the Bjorken variables of instanton-induced scattering 
processes\footnote{Our approach, focussing on the $I$-induced final state, 
differs substantially from an exploratory paper~\cite{bb} on the 
$I$-contribution to the (inclusive) parton structure functions. 
Ref.~\cite{bb} involves implicit integrations over the Bjorken variables of 
the $I$-induced scattering process. Unlike our approach, the calculations 
in Ref.~\cite{bb} are therefore bound to break down in the
interesting domain of smaller $x_{\rm Bj}{\mbox{\,\raisebox{.3ex}
    {$<$}$\!\!\!\!\!$\raisebox{-.9ex}{$\sim$}\,}} 0.3$, where most
of the data are located.}
 allow us to stay within the range of validity of instanton 
perturbation theory, as inferred from the lattice~\cite{rs-pl,mrs2}. 
We review the basic theoretical inputs to QCDINS, 
a Monte Carlo generator for instanton-induced processes in deep-inelastic 
scattering~\cite{grs,rs-qcdins}. Finally, we discuss the final state
characteristics of instanton-induced events.  

\begin{figure}
\begin{center}
\includegraphics[width=.7\textwidth]{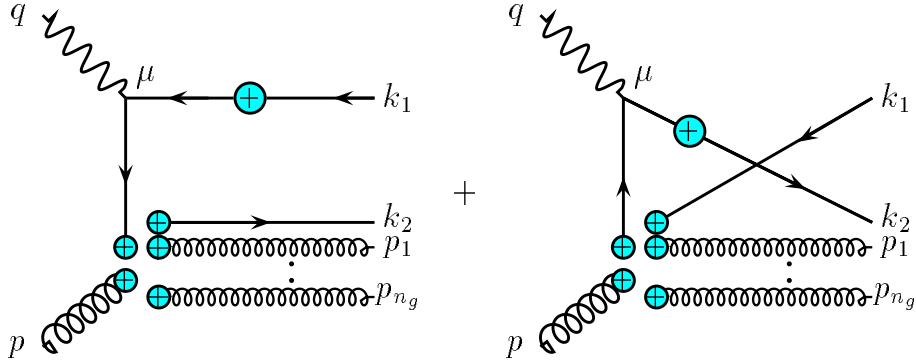}
\end{center}
\caption[]{
  Instanton-induced chirality-violating process, $\gamma^{\ast}(q)
  +g(p )\rightarrow \overline{{q}}_{R}(k_{1})+{
    q}_{R}(k_{2})+g(p_1)+\ldots +g(p_{n_g})$, for $n_f=1$ in leading order of 
$I$-perturbation theory. The corresponding
  Green's function involves the products of the appropriate classical fields
  (lines ending at blobs) as well as the (non-zero mode) quark propagator in 
the instanton
  background (quark line with central blob).  }
\label{ampl_ld}
\end{figure}

Let us  consider a generic $I$-induced process in deep-inelastic 
scattering (DIS),    
\begin{eqnarray}
      \gamma^\ast + g \Rightarrow 
      \sum_{\rm flavours}^{n_{f}}\left[\overline{q}_R+ q_R\right] 
      + n_g\,g\, ,
\end{eqnarray}
which violates chirality according to the anomaly~(\ref{anomaly}).
The corresponding scattering amplitude is calculated as follows~\cite{mrs1}:
The respective Green's function is first
set up according to instanton perturbation theory in Euclidean 
position space, then Fourier transformed to 
momentum space, LSZ amputated, and finally continued to Minkowski space
where the actual on-shell limits are taken. 
Again, the amplitude appears in the form of of an integral over
the collective coordinates~\cite{mrs1},  
\begin{eqnarray}
      {\mathcal T}_\mu ^{{(I)}\,(2n_f+n_g)}\,
      =
      \int\limits_0^\infty {d\rho}\, D ({\rho} )\int dU\,
      {\mathcal A}^{{(I)}\,(2n_f+n_g)}_\mu ({\rho},U ) \,.
\label{ampl-pt}
\end{eqnarray}
In leading order, the momentum dependence of the amplitude for fixed $\rho$ 
and $U$,  
\begin{eqnarray}
      {\mathcal A}^{{(I)}\,(2n_f+n_g)}_\mu 
      (q,p;k_1,k_{2},\ldots, k_{2n_f},p_1,\ldots , p_{n_g};{\rho},U) 
\, ,
\end{eqnarray}
factorizes, as illustrated in Fig.~\ref{ampl_ld} for the case $n_f=1$:  
The amplitude decomposes into a product of Fourier transforms 
of classical fields (instanton gauge fields; quark zero modes, e.g. as in 
Eq.~(\ref{twopointan})) and effective 
photon-quark ``vertices'' ${{\mathcal V}_\mu^{(t(u))}}(q,-k_{1(2)};\rho , U)$,
involving the (non-zero mode) quark propagator~\cite{brown} in the instanton 
background. 
These vertices are most important in the following argumentation since they 
are the only place where the space-like virtuality $-q^2=Q^2>0$ of the
photon enters. 

After a long and tedious calculation one finds~\cite{mrs1} for these vertices, 
\begin{eqnarray}
{\mathcal V}_\mu^{(t)}
(q,-k_1; \rho , U )&=&2\pi{\rm i}\rho^{3/2}
\left[ \epsilon \sigma_\mu  
{\overline{V} (q,k_1;\rho )} U^\dagger \right] \, ,
\\
{\mathcal V}_\mu^{(u)}
(q ,-k_2; \rho , U )&=&2\pi {\rm i} \rho^{3/2}\,
\left[ U  { V (q,k_2;\rho )} 
\overline{\sigma}_\mu \epsilon \right] \, ,
\end{eqnarray}
where
\begin{eqnarray}
\label{V}
 { V (q,k;\rho )} &=&
\left[  
\frac{\left( q-k\right)}{-(q-k)^2}
+\frac{k}{2 q\cdot k}
\right]{\rho\sqrt{-\left( q-k\right)^2}\,
K_1\left(\rho\sqrt{-\left( q-k\right)^2}\right)}
\\ \nonumber 
&& 
-
\frac{k}{2 q\cdot k}{\rho\sqrt{-q^{2}}\,
K_1\left(\rho\sqrt{-q^{2}}\right)} 
 \, .
\end{eqnarray}
Here comes the crucial observation: 
Due to the (large) space-like virtualities $Q^2=-q^2>0$ and $Q^{\prime 2}=
-(q-k)^2\geq 0$  in DIS and  
the exponential decrease of the Bessel $K$-function for large arguments
in Eq.~(\ref{V}), the $I$-size integration in our perturbative expression 
(\ref{ampl-pt}) for the amplitude is effectively cut off.     
Only small size instantons, $\rho \sim 1/\mathcal{Q}$, are probed 
in DIS and the predictivity of $I$-perturbation theory is retained for
sufficiently large ${\mathcal Q}={\rm min}(Q,Q^\prime)$.

\begin{figure}
\begin{center}
\includegraphics[width=.8\textwidth]{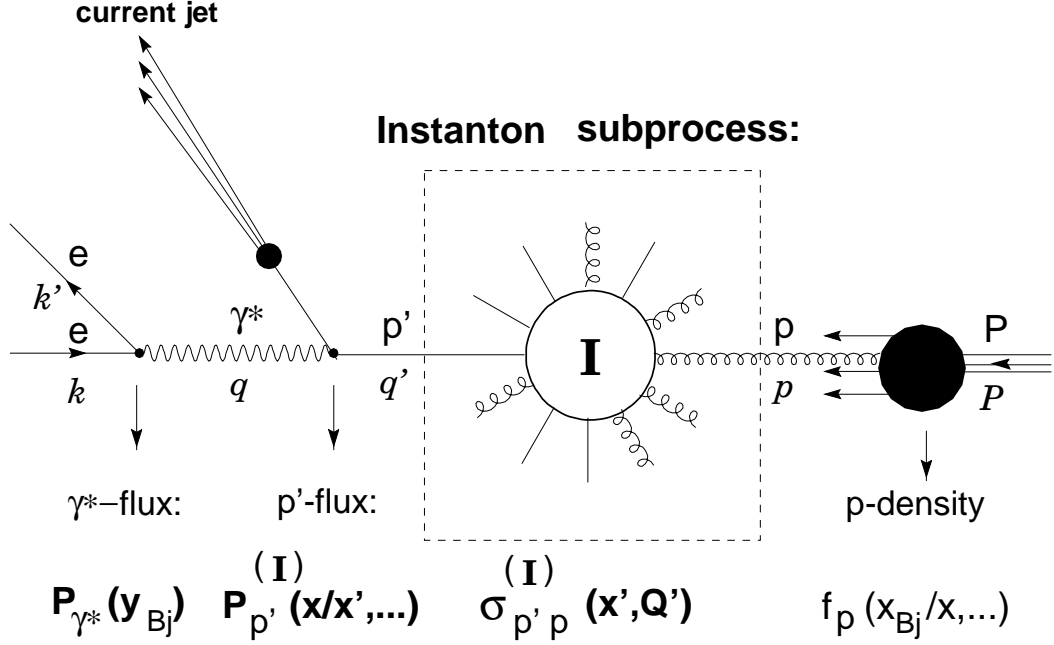}
\end{center}
\caption[]{
The leading instanton-induced process in the deep-inelastic regime of 
$e^\pm$\,P scattering ($n_f=3$).}
\label{ev-displ}
\end{figure}

The leading\footnote{$I$-induced processes initiated by a quark from the 
proton 
are suppressed by a factor of $\alpha_s^2$ with respect to the gluon initiated
process~\cite{rs-pl}. This fact, together with 
the high gluon density in the relevant kinematical domain at HERA,
justifies to neglect  quark initiated processes.} instanton-induced process 
in the DIS regime of $e^\pm P$ scattering for large photon virtuality $Q^2$ 
is illustrated in Fig.~\ref{ev-displ}. 
The inclusive $I$-induced cross section can be expressed as a 
convolution~\cite{rs,rs-pl}, involving 
integrations over the target-gluon density,
$f_g$, the virtual photon flux,
$P_{\gamma^\ast}$, and the known~\cite{rs-pl,mrs2} 
flux $P^{(I)}_{q^\prime}$ of the virtual quark 
$q^\prime$ in the $I$-background (c.f. Fig.~\ref{ev-displ}).

The crucial instanton-dynamics resides in the 
so-called instanton-subprocess (c.f. dashed box in Fig.~\ref{ev-displ})
with its associated total cross section 
$\sigma_{q^\prime g}^{(I)}(Q^\prime,x^\prime)$, depending on 
its own Bjorken variables,
\begin{equation}
Q^{\prime\,2}= -q^{\prime\,2}\ge 0;\hspace{2ex}
x^\prime =\frac{Q^{\prime\,2}}{2 p\cdot  q^\prime}\le 1\, .
\end{equation}
The cross section is obtained~\cite{rs-pl,mrs2} in the 
form of an integral over $I\overline{I}$ collective coordinates\footnote{Both
an instanton and an anti-instanton enter here, 
since cross sections result from taking the modulus squared of an
amplitude in the single $I$-background. In the present context, the 
$I\overline{I}$-interaction 
$\Omega$ takes into account the exponentiation of final state 
gluons~\cite{rs-pl}.},
\begin{eqnarray}
\nonumber
\sigma_{q^\prime g}^{(I)} &\sim & 
      \int d^4 R
      \int\limits_0^\infty d\rho 
      \int\limits_0^\infty d\overline{\rho}\, D(\rho) D(\overline{\rho})
       \int dU
            {\rm e}^{{-\frac{4\pi}{\alpha_s}}
      \Omega\left(\frac{R^2}{\rho\overline{\rho}},
      \frac{\overline{\rho}}{\rho},U \right)}
      \omega\left(\frac{R^2}{\rho\overline{\rho}},
      \frac{\overline{\rho}}{\rho},U \right)^{2n_f-1}
\\  && \times\ 
             {\rm e}^{-Q^\prime(\rho+\overline{\rho})}
      \,{\rm e}^{{\rm i} (p+q^\prime)\cdot R}\ \{\ldots \}
\, .
\label{cs}
\end{eqnarray}
Thus, as anticipated in Sect.~\ref{I-vacuum}, the group averaged  distribution 
of $I\overline{I}$-pairs (\ref{IaI}) is closely related to the 
instanton-induced cross section. The lattice constraints on this quantity 
are therefore extremely useful.  

\begin{figure}
\begin{center}
\includegraphics[angle=-90,width=.47\textwidth]{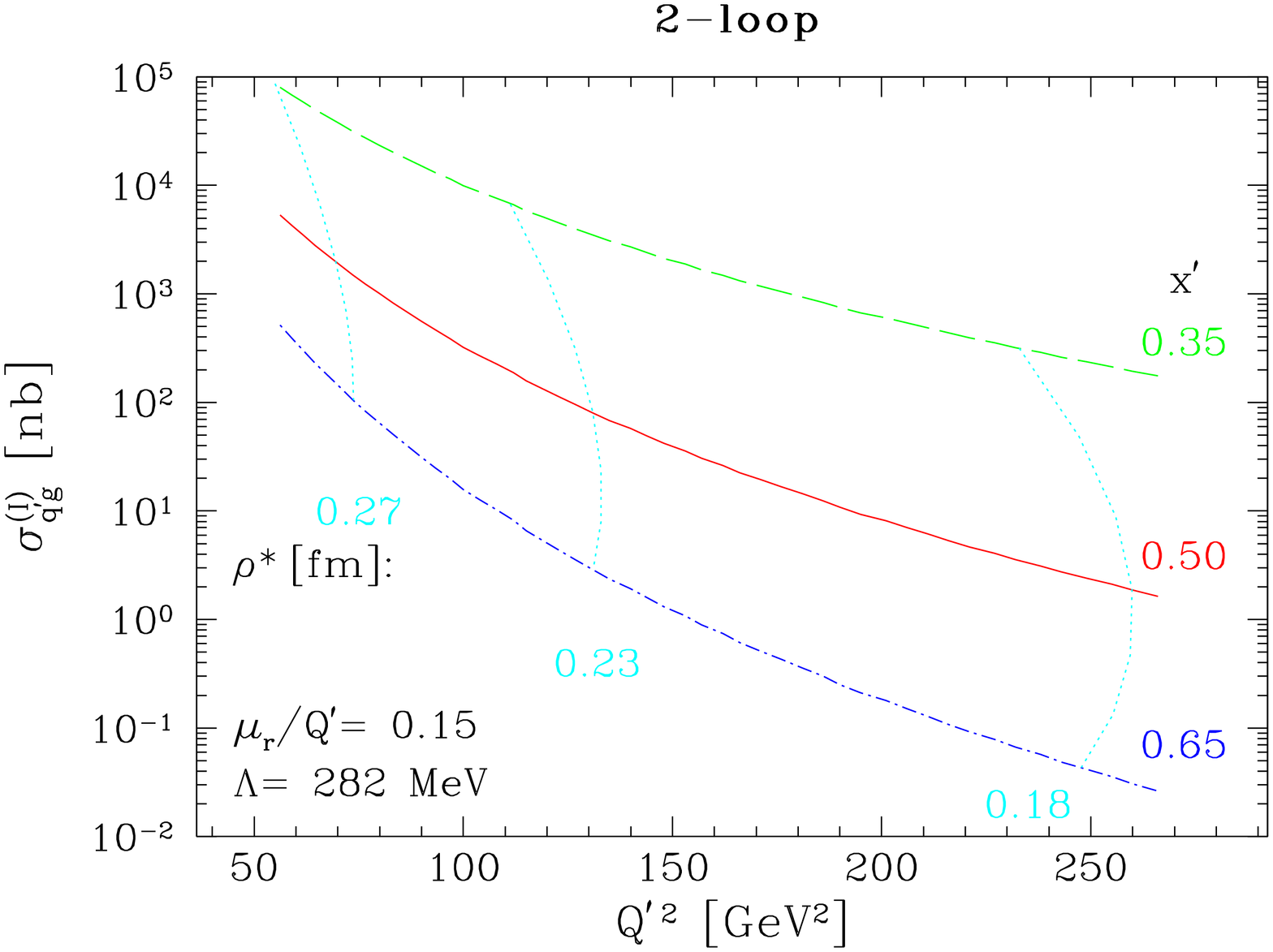}
\hspace{.02\textwidth}
\includegraphics[angle=-90,width=.47\textwidth]{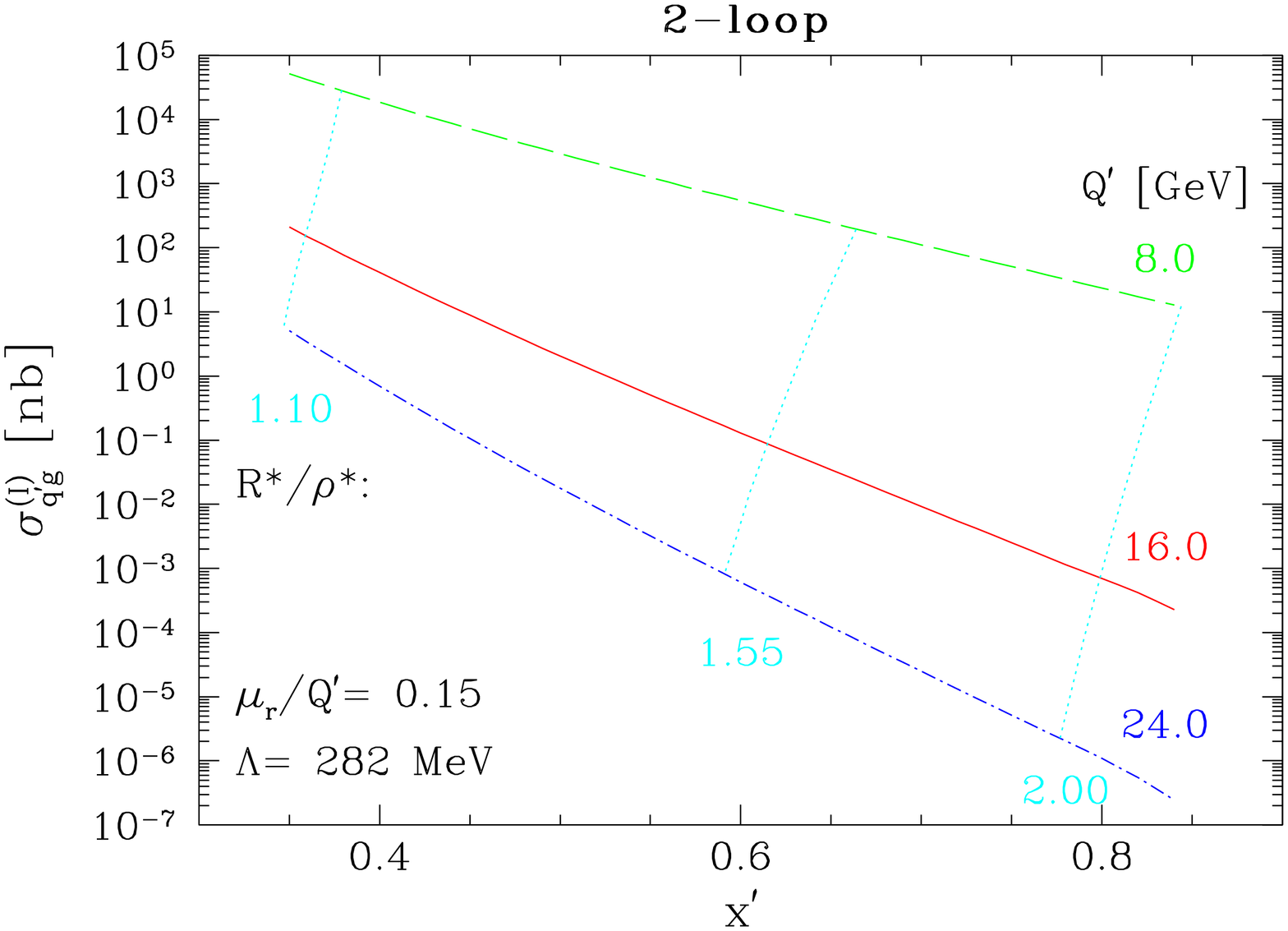}
\end{center}
\caption[]
   {$I$-subprocess cross section~\cite{rs-pl} displayed versus 
   the Bjorken variable $Q^{\prime\,2}$ with
   $x^\prime$ fixed
   (left) and versus $x^\prime$ with $Q^{\prime\,2}$ fixed (right) for 
    $n_f=3$. The dotted lines indicate the corresponding effective
    $I$-sizes $\rho^\ast$ [fm] (left) and $I\overline{I}$-distances
    $R^\ast$ in units of $\rho^\ast$ (right), respectively.}
\label{isorho}
\end{figure}

Again, the quark virtuality $Q^{\prime 2}$ cuts off large instantons.
Hence, the integrals in (\ref{cs}) are finite.
In fact, they are dominated by a unique saddle-point~\cite{rs-pl,mrs2},
\begin{eqnarray}
\nonumber
U^\ast &=& {\rm most\ attractive\ relative\ orientation}\, ;
\\ 
\label{saddle}
\rho^\ast& =& \overline{\rho}^\ast\sim 1/Q^\prime;\hspace{0.5cm} 
R^{\ast 2}\sim 1/(p+q^\prime)^2\ \Rightarrow \ 
\frac{R^\ast}{\rho^\ast}\sim \sqrt{\frac{x^\prime}{1-x^\prime}},
\end{eqnarray}
from which it becomes apparent (c.f. Fig.~\ref{isorho}) that the virtuality 
$Q^\prime$ contr{\^o}ls the
effective $I$-size, while $x^\prime$ determines the effective
$I\overline{I}$-distance (in units of the size $\rho$).
By means of the discussed 
saddle-point correspondence (\ref{saddle}), the lattice constraints
may be converted into a ``fiducial'' region for our 
cross section predictions in DIS~\cite{rs-pl},
\begin{equation}
 \left.\begin{array}{lcl}\rho^\ast& {\mbox{\,\raisebox{.3ex}
    {$<$}$\!\!\!\!\!$\raisebox{-.9ex}{$\sim$}\,}}&  
         0.3-0.35 {\rm\ fm};\\[1ex]
 \frac{R^\ast}{\rho^\ast}&{\mbox{\,\raisebox{.3ex}
    {$>$}$\!\!\!\!\!$\raisebox{-.9ex}{$\sim$}}\,}& 1\\
 \end{array}\right\}\Rightarrow
 \left\{\begin{array}{lcl}Q^\prime/\Lambda^{(n_f)}_{\overline{\rm MS}} 
&{\mbox{\,\raisebox{.3ex}
    {$>$}$\!\!\!\!\!$\raisebox{-.9ex}{$\sim$}}\,} &
 30.8;\\[1ex]
 x^\prime&{\mbox{\,\raisebox{.3ex}
    {$>$}$\!\!\!\!\!$\raisebox{-.9ex}{$\sim$}}\,} &0.35.\\
 \end{array} \right .
\label{fiducial}
\end{equation}
As illustrated in Fig.~\ref{isorho}, 
$\sigma_{q^\prime g}^{(I)}(Q^\prime,x^\prime)$ 
 is very steeply growing for decreasing values of $Q^{\prime 2}$ and 
$x^\prime$, respectively. The constraints~(\ref{fiducial}) from lattice 
simulations are extremely valuable for making concrete predictions.   
Note that the fiducial region~(\ref{fiducial}) and thus all our predictions
for HERA never involve values of the $I\overline{I}$-interaction 
$\Omega$ smaller than $-0.5$ (c.f. Fig.~\ref{omega}), 
a value often advocated as a lower reliability bound~\cite{hgbound}.

Let us present an update of our published prediction~\cite{rs-pl} of the 
$I$-induced cross section at HERA. For the following \emph{modified} 
standard cuts, 
\begin{eqnarray}
\label{standard-cuts} 
{\mathcal C}_{\rm std}&=& 
x^\prime\ge0.35,\ Q^\prime\ge 30.8\,\Lambda^{(n_f)}_{\overline{\rm MS}},\  
x_{\rm Bj}\geq 10^{-3},
\\ \nonumber &&
0.1\leq y_{\rm Bj}\leq 0.9 ,\ Q \ge 
30.8\,\Lambda^{(n_f)}_{\overline{\rm MS}}, 
\end{eqnarray}
involving the minimal cuts~(\ref{fiducial}) extracted from lattice 
simulations, and an update of $\Lambda_{\overline{\rm MS}}$ to the 
1998 world average~\cite{pdg}, we obtain 
\begin{eqnarray}
\label{minimal-cuts}
\sigma^{(I)}_{\rm HERA}({\mathcal C}_{\rm std}) 
&=& 29.2^{+9.9}_{-8.1}\, {\rm pb}.
\end{eqnarray}
Note that the quoted errors in the cross section~(\ref{minimal-cuts})
only reflect the uncertainty in 
$\Lambda_{\overline{\rm MS}}^{(5)}=219^{+25}_{-23}$ MeV~\cite{pdg},
on which $\sigma^{(I)}$ is known to depend very strongly~\cite{rs-pl}.
We have also used now the 3-loop formalism~\cite{pdg} to perform
the flavour reduction of $\Lambda_{\overline{\rm MS}}^{(n_f)}$ from
5 to 3 light flavours. Finally, the value of $\sigma^{(I)}$ is substantially
reduced compared to the one in Ref.~\cite{rs-pl}, since we preferred 
to introduce a further cut in $Q^2$, with 
$Q^2_{\rm min} =Q^{\prime 2}_{\rm min}$, in order to insure the smallness 
of the $I$-size $\rho$ in contributions associated with the second term in 
Eq.~(\ref{V}).     

\begin{figure}[h]
\begin{center}
\includegraphics[width=.6\textwidth]{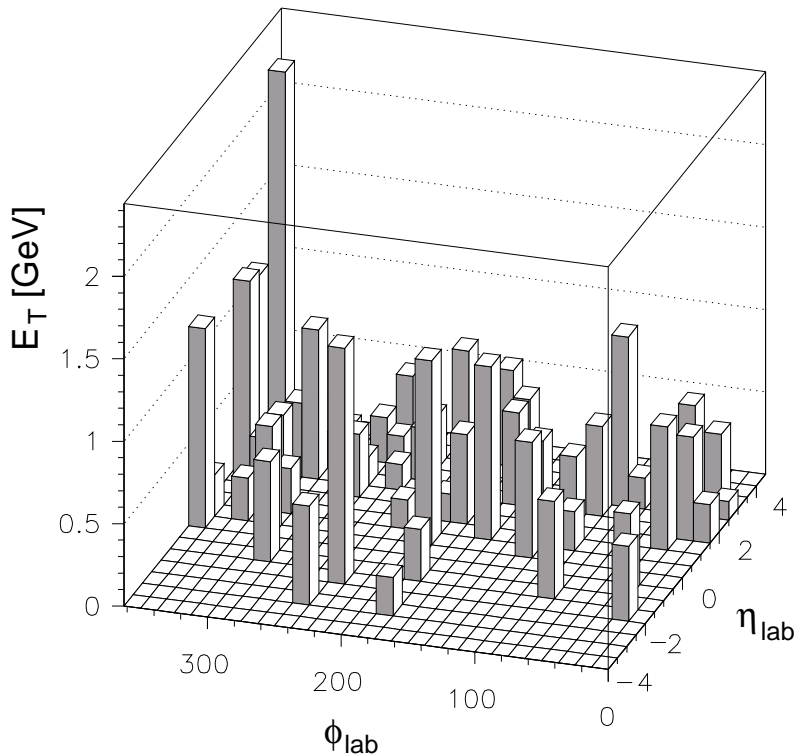}
\end{center}
\caption[]
     {Lego plot of a typical instanton-induced event from QCDINS.}
\label{lego}
\end{figure}

Based on the predictions of $I$-perturbation theory, a Monte Carlo generator 
for simulating QCD-instanton induced scattering processes in DIS, 
QCDINS, has been developed~\cite{grs,rs-qcdins}.   
It is designed as an 
``add-on'' hard process generator interfaced by default to the Monte
Carlo generator HERWIG~\cite{herwig}. Optionally, an interface to 
JETSET~\cite{jetset} is also available for the final hadronization step. 

QCDINS incorporates the essential characteristics that have been derived 
theoretically for the hadronic final state of $I$-induced processes:
notably, the isotropic production of the partonic final
state in the $I$-rest system ($q^\prime g$ center of mass system in 
Fig.~\ref{ev-displ}), 
flavour ``democracy'', energy
weight factors different for gluons and quarks, and  a high average
multiplicity $2n_f+{\cal O}(1/\alpha_s)$ of produced partons with a 
(approximate) Poisson distribution of the gluon multiplicity.  

The characteristic features of the $I$-induced final state are
illustrated in Fig.~\ref{lego} displaying the lego plot of a
typical event from QCDINS (c.\,f. also Fig.~\ref{ev-displ}):
Besides a single (not very hard) current jet, one expects  an
accompanying densely populated ``hadronic band''. 
For $x_{\rm Bj\,min}\simeq
10^{-3}$, say,  it is centered around $ \overline{\eta} \simeq 2$  and
has a width of $\Delta \eta \simeq \pm 1$. The band directly reflects the 
isotropic production of an $I$-induced ``fireball'' of ${\cal
O}(10)$  partons in the $I$-rest system. Both the total transverse
energy $\langle E_T \rangle \simeq 15$ GeV and the charged particle
multiplicity $\langle n_c\rangle \simeq 13$ in the band are far higher than in
normal DIS events. Finally, each $I$-induced event has to contain
strangeness 
such that the number of $K^0$'s amounts to $\simeq 2.2$/event. 

\begin{figure}
\begin{center}
\includegraphics[width=.9\textwidth]{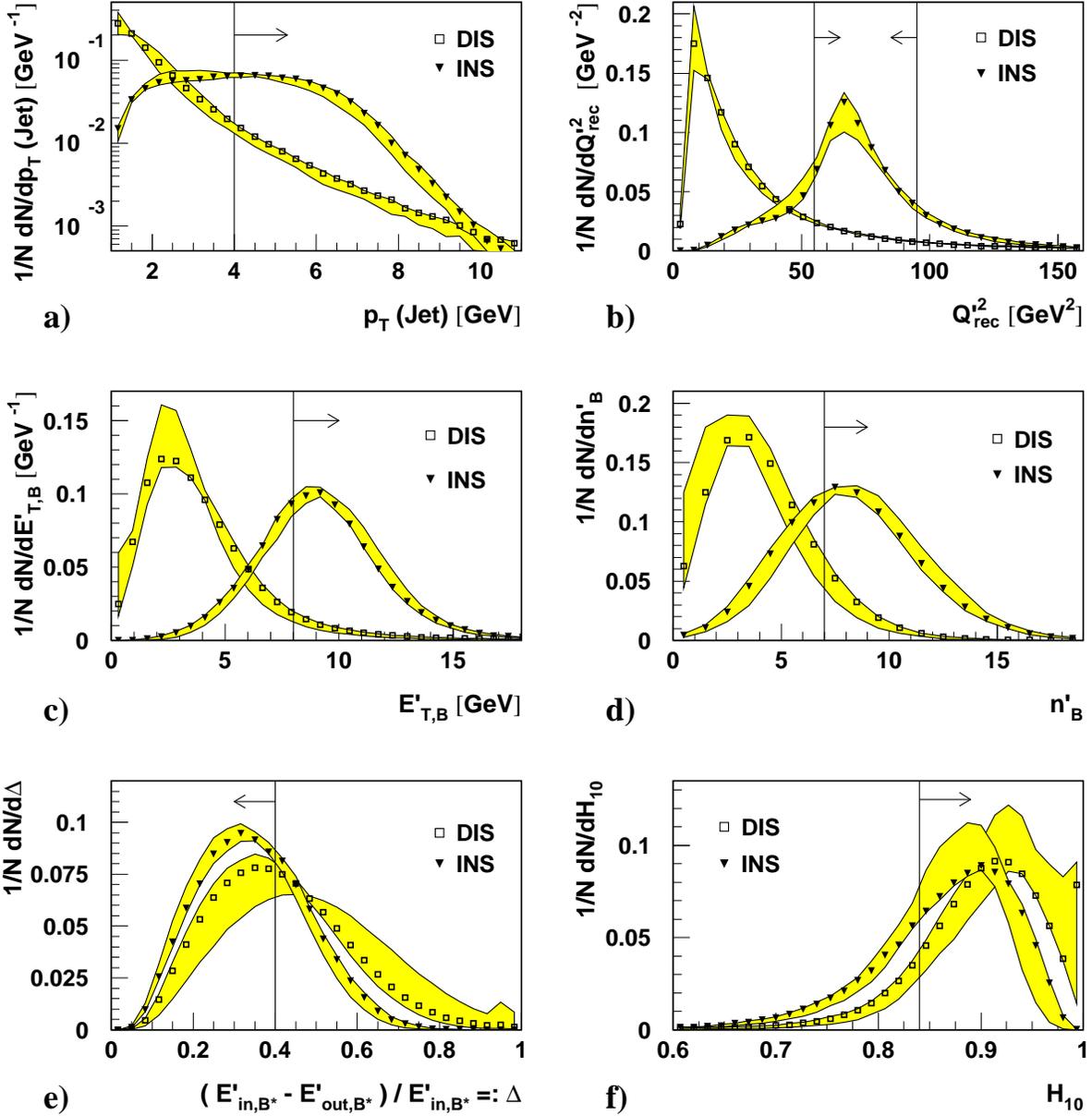}
\end{center}
\caption[]
    {Distributions of various observables for normal DIS and 
    $I$-induced processes~\cite{cgrs}. Shown are the distributions
    for the ``reference Monte Carlos'' (INS markers = QCDINS + HERWIG,
    DIS markers = ARIADNE~\cite{ariadne}, including Pomeron exchange) and 
    their variations
    (shaded band) resulting from the choice of different models or
    the variation of parameters of a model
    (c.\,f. Fig.~\ref{var-signal}). The lines and the corresponding arrows 
    show 
    the cut applied in each of the observables, with the arrows pointing
    in the direction of the allowed region.}
\label{6var}
\end{figure}

\section{\label{search}Search Strategies}

In a recent detailed study~\cite{cgrs}, based on QCDINS and standard DIS 
event generators, a number of basic (experimental) questions have been
investigated:  How to  isolate an $I$-enriched data sample by means of cuts 
to a set of observables?
How large are the dependencies on Monte-Carlo models, both for $I$-induced
(INS) and normal DIS events? Can the Bjorken-variables 
$(Q^\prime,\ x^\prime)$ of the $I$-subprocess be reconstructed? 

All the studies presented in Ref.~\cite{cgrs} were performed in the hadronic 
center of mass frame, which is a suitable frame of reference
in view of a good distinction between
$I$-induced and normal DIS events (c.\,f. Ref.~\cite{jgerigk}).
The results are based on a study of the hadronic final
state, with typical acceptance cuts of a HERA detector being applied. 

Let us briefly summarize the main results of Ref.~\cite{cgrs}. 
While the ``$I$-separation power''= $\rm INS_{\rm
eff(iciency)}/DIS_{\rm eff(iciency)}$ typically does not exceed ${\cal O}(20)$ 
for single observable cuts, a set of six observables 
(among $\sim 30$ investigated in Ref.~\cite{jgerigk})) with much improved
joint $I$-separation power $={\cal O}(130)$  could be found, see 
Fig.~\ref{6var}. These are (a) the $p_T$ of the current jet, 
(b) $Q^{\prime 2}$ as reconstructed from the final state, (c) 
the transverse energy and (d) the number of charged particles in the
$I$-band region\footnote{With the prime in Fig.~\ref{6var} (c,d,e) indicating 
that the hadrons from the current 
jet have been subtracted.}, and (e,f) two shape observables that are sensitive 
to the event isotropy.    

\begin{figure}
\begin{center}
\includegraphics[width=.95\textwidth]{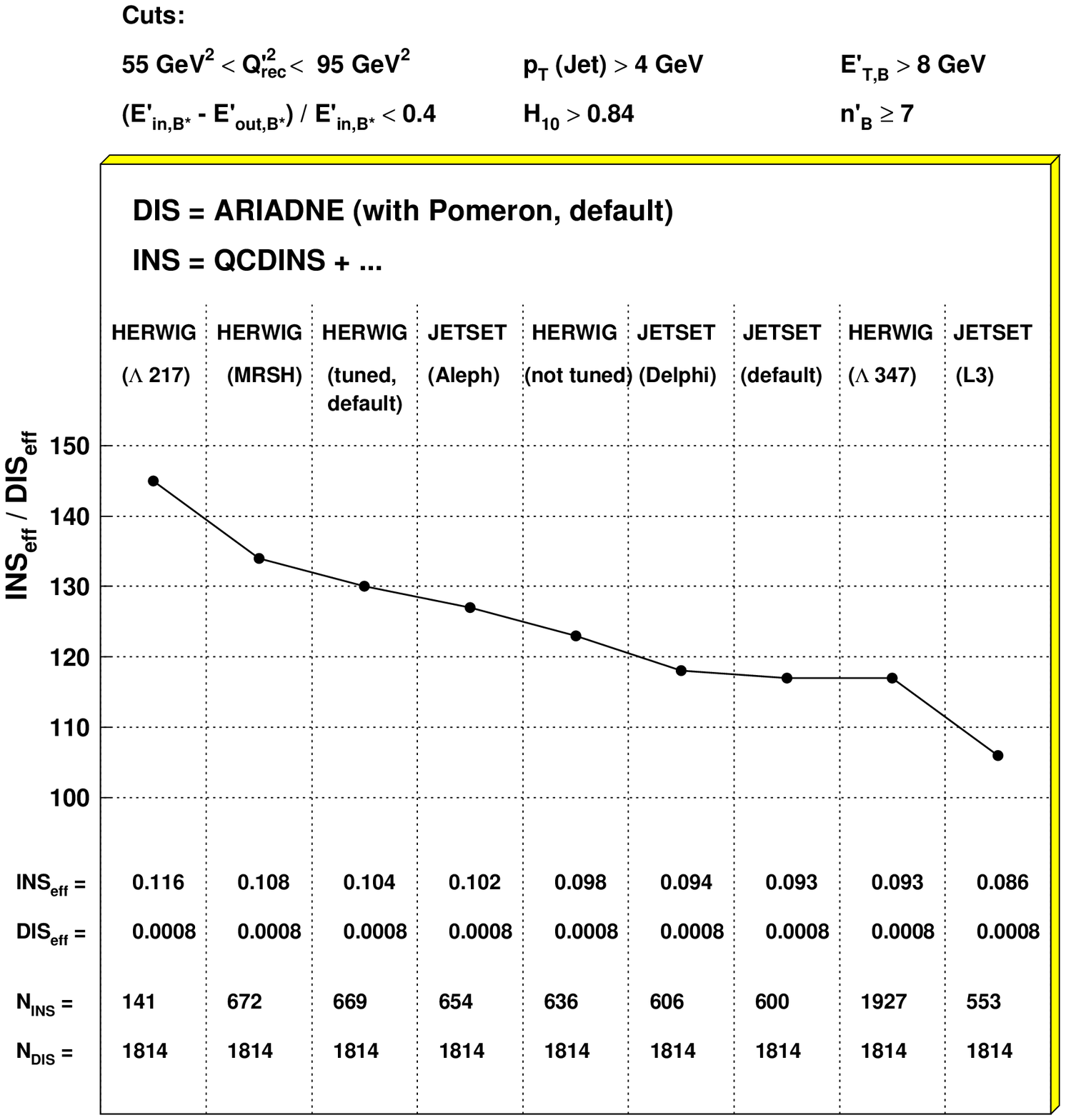}
\end{center}
\caption[]
    {$I$-separation power (INS$_{\rm eff}$/DIS$_{\rm eff}$) 
    of a multidimensional cut-scenario depending on
    the variation of MC models and parameters used to simulate
    $I$-induced events~\cite{cgrs}. The efficiencies 
    and remaining event numbers for an integrated luminosity 
    $\mathcal{L} \simeq 30~\mbox{pb}^{-1}$ and corresponding to the cross 
    section from
    QCDINS 1.6.0 are listed.}
\label{var-signal}
\end{figure}

The systematics
induced by varying the modelling of $I$-induced events remains
surprisingly small (Fig.~\ref{var-signal}). In contrast,   
the modelling of normal DIS events in the relevant region of phase
space turns out to depend quite strongly on the used generators and
parameters~\cite{cgrs}. Despite a relatively high
expected rate for $I$-events in the fiducial 
DIS region~\cite{rs-pl}, a better understanding of the tails of
distributions for normal DIS events turns out to be quite important.

\vfill\eject


\begin{thebibliography}{8.}
\addcontentsline{toc}{section}{References}

\bibitem{bpst}
A. Belavin, A. Polyakov, A. Schwarz, Yu. Tyupkin:  
Phys. Lett. B \textbf{59}, {85} {(1975)}
\bibitem{th} 
G. `t Hooft: Phys. Rev. Lett. \textbf{37}, {8} {(1976)};
Phys. Rev. D \textbf{14}, {3432} {(1976)}; 
Phys. Rev. D \textbf{18}, {2199} {(1978)}
(Erratum); Phys. Rep. \textbf{142}, {357} {(1986)}
\bibitem{ssh}
T. Sch{\"a}fer, E. Shuryak:
Rev. Mod. Phys. \textbf{70}, 323 (1998)
\bibitem{rs}
A. Ringwald, F. Schrempp: `Towards the Phenomenology 
of QCD-Instanton Induced Particle Production at HERA',
hep-ph/9411217. 
In:
\emph{Quarks `94, Proc. 8th Int. Seminar, Vladimir, Russia, May 11--18, 1994}, 
ed. by D. Grigoriev et al. (World Scientific, Singapore 1995) pp. 170--193
\bibitem{grs}
M. Gibbs, A. Ringwald, F. Schrempp: 
`QCD-Instanton Induced Final States in Deep Inelastic Scattering', 
hep-ph/9506392. 
In: 
\emph{Workshop on Deep Inelastic Scattering and QCD (DIS 95), Paris, France, 
April 24--28, 1995},
ed. by J.F. Laporte, Y. Sirois (Ecole Polytechnique, Paris 1995) 
pp. 341-344
\bibitem{jgerigk}
J. Gerigk: `QCD-Instanton-induzierte Prozesse in tiefunelastischer
$e^\pm p$-Streuung', Dipl. Thesis (in German), University of Hamburg 
(unpublished) and MPI-PhE/98-20, Nov. 1998
\bibitem{cgrs}
T. Carli, J. Gerigk, A. Ringwald, F. Schrempp: 
`QCD Instanton-Induced Processes in Deep-Inelastic Scattering --
Search Stragegies and Model Dependencies', hep-ph/9906441.
To appear in: \emph{Proc. DESY Workshop 1998/1999 on Monte Carlo
Generators for HERA Physics} 
\bibitem{mrs1}
S. Moch, A. Ringwald, F. Schrempp: Nucl. Phys. B \textbf{507}, {134} {(1997)}
\bibitem{rs-pl}
A. Ringwald, F. Schrempp: Phys. Lett. B \textbf{438}, {217} {(1998)}
\bibitem{mrs2}
S. Moch, A. Ringwald, F. Schrempp: in preparation
\bibitem{rs-lat}
A. Ringwald, F. Schrempp:  
Phys. Lett. B \textbf{459}, 249 (1999) 
\bibitem{ukqcd}
D.A. Smith, M.J. Teper (UKQCD): Phys. Rev. D 
\textbf{58}, {014505} (1998)
\bibitem{rs-qcdins}
A. Ringwald, F. Schrempp: in preparation
\bibitem{ber}
C. Bernard: Phys. Rev. D \textbf{19}, 3013 (1979)
\bibitem{morretal}
T. Morris, D. Ross, C. Sachrajda: 
Nucl. Phys. B \textbf{255}, 115 (1985)
\bibitem{dMS}
A. Hasenfratz, P. Hasenfratz: Nucl. Phys. B \textbf{193}, 210 (1981)
\hfill\break
M. L{\"u}scher: Nucl. Phys. B \textbf{205}, 483 (1982)
\bibitem{chu}
M.-C. Chu, J.M. Grandy, S. Huang, J.W. Negele: 
Phys. Rev. D \textbf{49}, 6039 (1994)
\bibitem{mike}
M. Teper: private communication
\bibitem{alpha}
S. Capitani, M. L{\"u}scher, R. Sommer, H. Wittig: 
Nucl. Phys. B \textbf{544}, {669} {(1999)}
\bibitem{anomaly}
S. Adler: Phys. Rev. \textbf{177}, 2426 (1969)
\hfill\break
J. Bell, R. Jackiw: Nuovo Cimento \textbf{51}, 47 (1969)
\hfill\break
W. Bardeen: Phys. Rev. \textbf{184}, 1848 (1969)
\bibitem{atiyah}
M. Atiyah, I. Singer: Ann. Math. \textbf{87}, 484 (1968)
\bibitem{yung}
A. Yung:
Nucl. Phys. B \textbf{297}, 47 (1988)
\bibitem{valley-most-attr-orient}
V.V. Khoze, A. Ringwald: 
Phys. Lett. B \textbf{259}, 106 (1991)
\bibitem{valley-gen-orient}
J. Verbaarschot: 
Nucl. Phys. B \textbf{362}, 33 (1991)
\bibitem{bb}
I. Balitsky, V. Braun: Phys. Lett. B \textbf{314}, 237 (1993) 
\bibitem{brown}
L. Brown, R. Carlitz, D. Creamer, C. Lee:
Phys. Rev. D \textbf{17}, 1583 (1978)
\bibitem{hgbound}
V. Zakharov: Nucl. Phys. B \textbf{353}, 683 (1991)
\hfill\break
M. Maggiore, M. Shifman: Nucl. Phys. B \textbf{365}, 161 (1991); 
\emph{ibid. } \textbf{371}, 177 (1991)
\hfill\break
G. Veneziano: Mod. Phys. Lett. A \textbf{7}, 1661 (1992)
\bibitem{pdg}
C. Caso et al. (Particle Data Group):
Eur. Phys. J. C \textbf{3}, 1 (1998)
\bibitem{herwig} 
G. Marchesini et al.: Comp. Phys. Commun. \textbf{67}, 465 (1992)
\bibitem{jetset}
T. Sj{\"o}strand: Comp. Phys. Commun. \textbf{82}, 74 (1994)
\bibitem{ariadne}
L. L{\"o}nnblad: Comp. Phys. Commun. \textbf{71}, 15 (1992)

\end{thebibliography}
\end{document}